\documentclass[pre,twocolumn,letterpaper,showpacs]{revtex4}

\usepackage{graphicx}

\begin{document}

\title{Experimental evidence of shock mitigation in a Hertzian tapered chain}

\author{Francisco Melo$^1$, St\'ephane Job$^2$, Francisco Santibanez$^1$, Franco Tapia$^1$}
\affiliation{
$(1)$ Departamento de F\'{\i}sica, Universidad de Santiago de Chile,\\
and Center for Advanced Interdisciplinary Research in Materials (CIMAT),\\
Av. Ecuador 3493, Casilla 307, Correo 2, Santiago de Chile.\\
$(2)$ {\sc Supmeca}, 3 rue Fernand Hainaut, 93407 Saint-Ouen Cedex, France.}

\date{\today}

\begin{abstract}
We present an experimental study of the mechanical impulse propagation through a horizontal alignment of elastic spheres of progressively decreasing diameter $\phi_n$, namely a tapered chain. Experimentally, the diameters of spheres which interact via the Hertz potential are selected to keep as close as possible to an exponential decrease, $\phi_{n+1}=(1-q)\phi_n$, where the experimental tapering factor is either $q_1\simeq5.60$~\% or $q_2\simeq8.27$~\%. In agreement with recent numerical results, an impulse initiated in a monodisperse chain (a chain of identical beads) propagates without shape changes, and progressively transfer its energy and momentum to a propagating tail when it further travels in a tapered chain. As a result, the front pulse of this wave decreases in amplitude and accelerates. Both effects are satisfactorily described by the hard spheres approximation, and basically, the shock mitigation is due to partial transmissions, from one bead to the next, of momentum and energy of the front pulse. In addition when small dissipation is included, a better agreement with experiments is found. A close analysis of the loading part of the experimental pulses demonstrates that the front wave adopts itself a self similar solution as it propagates in the tapered chain. Finally, our results corroborate the capability of these chains to thermalize propagating impulses and thereby act as shock absorbing devices.
\end{abstract}

\pacs{83.80.Fg, 05.45.-a, 62.50.+p}
\maketitle

%--------------------------------------------------------------------%
%-----> I) Introduction
%--------------------------------------------------------------------%

\section{Introduction}

Granular materials are useful for shock protection, and cast iron shots are a well known example of system effectively used in damping of contact explosive loading \cite{Nesterenko2001}. Recent numerical results have demonstrated that new features can still be developed, such as impulse thermalization \cite{Sen2001,Sen2003,Doney2005,SenNaka2005} or shock confinement \cite{Hong2005} in one dimensional granular media. Indeed, the propagation of an impulse through a chain of monodisperse elastic spheres exhibits interesting nonlinear physics \cite{Nesterenko2001}. When the spheres in the chain barely touch one another, the energy of an impulse initiated at one end of the chain propagates as a solitary wave of well defined velocity and width, as described by pioneering work of Nesterenko \cite{Nesterenko1984,Lazaridi1985,Nesterenko1994,Nesterenko1995,Nesterenko2001}, and as confirmed later by other authors \cite{Coste1997,Hascoet1999,Mackay99,Job2005}. When small dissipative effects such as viscoelastic restitution or solid friction (mainly thwarted rotation between beads and friction between beads and walls of the setup) are taken into account the solitary wave only attenuates in amplitude and spreads \cite{Job2005,Poschel2001}. Under static loading, the chain still supports solitary wave solutions \cite{Nesterenko2001,Coste1997,Mackay99}, in addition to dispersive linear sound waves of weak amplitudes. In the presence of a gradient of static loading (e.g. a vertical chain under gravity), it has been proved that propagation of an impulse is controlled by dispersion, and extra dispersion introduces a coupling between quasisolitary and oscillatory propagating modes \cite{Sinkovits95,Sen96,Hong99,Hascoet02}. However, dispersion does not allow the possibility of distributing the energy of a solitary wave uniformly, neither throughout a finite length vertical chain \cite{Sen96} nor in a horizontal linear chain \cite{Mackay99}. In regard to energy propagation in the absence of dispersion, P{\"o}schel and Brillantov \cite{Poschel2001} considered a one dimensional chain of spheres with a restitution coefficient that was velocity independent. By keeping constant the size of the beads at both end of the chain they showed that the energy transmission is optimal when the mass distribution is an exponentially decreasing function. Similarly, Nesterenko {\it et al} have shown \cite{Nesterenko1984,Lazaridi1985,Nesterenko2001} that the momentum of a pulse is totally transmitted when it passes through a sharp decrease of bead's size, and is partially reflected in the reverse case.

Here, we investigate experimentally the effect of shock mitigation occurring in an unloaded chain of exponentially decreasing bead size. This effect, recently predicted by Sen {\it et al} \cite{Sen2001,Sen2003,Doney2005,SenNaka2005}, is due to the progressive delay on the transmission of a fraction of the energy of the pulse when propagating in a sequence of increasing stiffness contacts defined by the sequence of smaller beads. Consistently, excepting the reflection of the pulse at the rigid end of the chain, there is no detectable reflected wave propagating backward while the pulse propagates down toward the smaller extremity of the tapered chain. Experimental evidence of this effect have been observed by Nakagawa {\it et al} \cite{Nakagawa2004}, when an impulse is introduced in a tapered chain in which the beads are initially barely in contact and the chain is let to expand after the impulse propagation, the last beads being initially at some distance of a wall sensor. This configuration is suitable to measure the speed of beads by imaging techniques and the force due to the ejection of the last bead colliding with a wall sensor. In our experiment the beads always keep in contact which allows us to adopt a complementary point of view and focus on the shape and speed of the wave instead of that of beads. The main advantage is that the wave picture is better realized and energy transfer from the front pulse to a propagating tail is fully visualized. This article is organized as follow. In section II, we describe our main experimental considerations and present force measurements acquired in two different chains, whose tapering factor is either $q_1\simeq5.60$~\% or $q_2\simeq8.27$~\%. Similarly, the acceleration of the front pulse is characterized as it propagates toward the decreasing end of the chain. Effects of misalignments and tapering factor variations are investigated as well. The section III is devoted to brief summary of the main theoretical aspects and discussion. We show that the ballistic approximation in conjunction with the effect of small restitution coefficient is well suited to capture the acceleration and the decrease of force amplitude of a pulse traveling through the tapered chain.

%--------------------------------------------------------------------%
%-----> II) Experimental observations
%--------------------------------------------------------------------%

\section{Experimental observations}

Beads are high carbon chrome hardened {\em AISI 52100} steel roll bearing. Density is $\rho=7780$~kg/m$^3$, Young's modulus is $Y=203\pm4$~GPa, Poisson ratio is $\nu=0.3$, and yield stress is $\sigma_Y\simeq2$~GPa, \cite{Tsubaki}. We have carefully checked that bead's deformation always keeps elastic behavior in all our experiments. Assuming that the contact surface is a disk of area $A=\pi({\theta}RF)^{2/3}$ \cite{Landau1967}, we estimate that yield stress occurs for a compression force greater or equal to $F_Y\simeq470$~N, which corresponds to an overlap $\delta_Y\simeq11$~$\mu$m, for beads of $26$mm diameter. This threshold is greater for smaller beads.

A solitary wave is initiated in a monodisperse chain made of $n_0=16$ beads whose diameter is $\phi_0=26$~mm, by applying a short impact with constant velocity (over all experiments) of a smaller bead ($\phi_i=8$~mm) at one of its extremity \cite{Job2005}. This impulse then reaches a tapered chain composed of beads whose diameters $\phi_n$ progressively decrease. More precisely, two tapered chains are implemented, whose intermediate diameters, $\phi_n$, are given in the table~\ref{tab:beaddiameter}. Tapering factor is determined as $q=1-\langle(\phi_n/\phi_0)^{n-n_0}\rangle_{n>n_0}$. As everywhere in this document, errorbars refer to the unbiased sample standard deviation (confidence interval, $CI\simeq68$~\%). The first chain provides a tapering factor of $q_1=(5.60\pm0.67)$~\%, and the second one provides a tapering factor of $q_2=(8.27\pm0.31)$~\%.

\begin{table}[h]
\scriptsize
\begin{tabular}{|c||c|c|c|c|c|c|c|c|c|} \hline
Chain & bead {\it n}  & 01-16 & 17    & 18    & 19    & 20    & 21    & 22    & 23    \\ \hline
1     & $\phi_n$ [mm] & 26.00 & 24.00 & 23.00 & 22.00 & 21.00 & 20.00 & 19.00 & 18.00 \\ \hline
2     & $\phi_n$ [mm] & 26.00 & 24.00 & 22.00 & 20.00 & 18.00 & 16.65 & 16.00 & 14.00 \\ \hline
\hline
Chain & bead {\it n}  & 24    & 25    & 26    & 27    & 28    & 29    & 30    & -     \\ \hline
1     & $\phi_n$ [mm] & 16.65 & 16.00 & 15.00 & 14.00 & 13.00 & 12.00 & 11.00 & -     \\ \hline
2     & $\phi_n$ [mm] & 13.00 & 12.00 & 11.00 & 10.00 &  9.00 & -     & -     & -     \\ \hline
\end{tabular}
\caption{\label{tab:beaddiameter}Bead's diameters.}
\end{table}

Beads of the monodisperse chain are aligned on a rigid Plexiglas track over which beads are allowed to roll with little friction. Beads of the tapered chain are located on a special stepped track composed of a collection of short tracks of length nearly the beads diameter, designed in such a way that the tapered chain is automatically aligned on the axis of the monodisperse chain. The whole rigid structure serving for beads alignment is fixed on an optical table, which guarantees a misalignment, i.e., off center error, lower than $2$~\% between the smallest consecutive beads, and lower than $0.1$~\% for the biggest beads considered here. It appears that repeatability of measurements is not so much affected by misalignments if the beads are not allowed to move laterally. This effect is quantified further in the text by introducing a controlled off center displacement of the beads.

%--------------------------------------------------------------------%
%-----> FIGURE #1 OVER 6
%--------------------------------------------------------------------%

\begin{figure}[ht]
\includegraphics{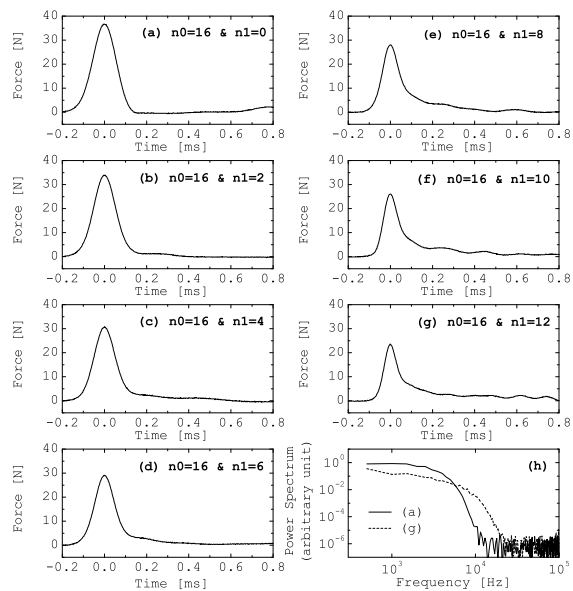}
\caption{\label{fig:Force_vs_position_q1}(a)-(g) Force as a function of time (arbitrary origin). Evolution of a pulse initiated in a monodisperse chain of $n_0=16$ beads, shown in (a), as it propagates in the tapered chain of $n_1$ beads with tapering factor $q=q_1$. (b)-(g) Actual forces felt at the right contact of beads \#$18$, $20$, $22$, $24$, $26$, and $28$ (see table~\ref{tab:beaddiameter}). (h) Power spectrum, with arbitrary unit, indicating a comparison between incoming and outcoming pulses in the tapered chain.}
\end{figure}

%--------------------------------------------------------------------%
%-----> FIGURE #2 OVER 6
%--------------------------------------------------------------------%

\begin{figure}[ht]
\includegraphics{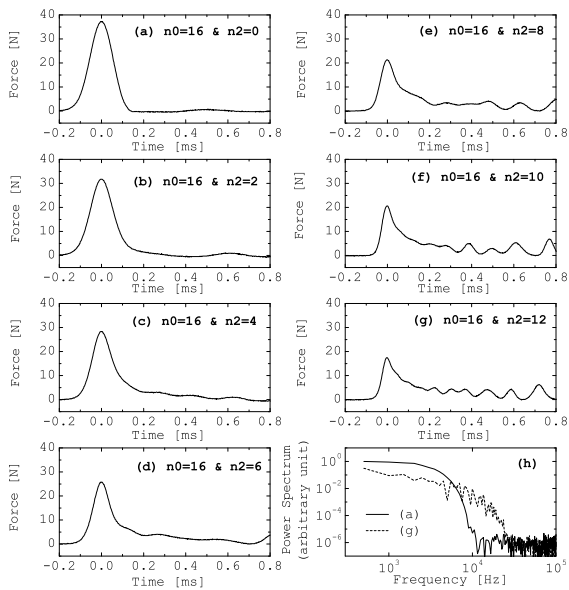}
\caption{\label{fig:Force_vs_position_q2}Same as Fig.~\ref{fig:Force_vs_position_q1}, for a pulse initiated in a monodisperse chain of $n_0=16$ beads, shown in (a), and propagating through the tapered chain of $n_2$ beads with tapering factor $q=q_2$. (b)-(g) Actual forces felt at the right contact of beads \#$18$, $20$, $22$, $24$, $26$, and $28$ (see table~\ref{tab:beaddiameter}). (h) Power spectrum, with arbitrary unit, indicating a comparison between incoming and outcoming pulses in the tapered chain.}
\end{figure}

A piezoelectric dynamic impulse sensor ({\em PCB 208A11} with sensitivity $112.40$~mV/N) located at the end of the chain provides the force at the rigid end. This sensor has a flat cap made of the same material as the beads. Forces inside the monodisperse chain are monitored by a flat dynamic impulse sensor ({\em PCB 200B02} with sensitivity $11.24$~mV/N) that is inserted inside one of the beads, cut in two parts. The total mass of the bead sensor system has been compensated to match the mass of an original bead. This system allows achieving non intrusive force measurement, by preserving both contact and inertial properties of the bead-sensor system. The stiffness of the sensor $k_s=1.9$~kN/$\mu$m being greater than the stiffness of the Hertzian contact ($k_s\gg k_{H}\propto \kappa\delta^{1/2}$, see section III for definitions of $\kappa$ and $\delta$), the coupling between the chain and the sensor is consequently negligible. Additional details on how to relate the force $F_s$ detected by the sensor with the actual force at the contact between two beads are given in Ref. \cite{Job2005}. Signals from sensors are amplified by a conditioner ({\em PCB
482A16}), recorded by a two channels numeric oscilloscope ({\em Tektronix TDS340}), and transferred to a computer. Measurements have been repeated nine times and averaged to minimize errors, and acquisitions are achieved as follows. In the monodisperse chain, the sensor at the end of the chain serves as a trigger, and the force is recorded at various positions by the sensor in the chain. In the tapered chain, force is recorded by the sensor at the end of the chain, while the sensor in the monodisperse chain is let at a given position to trig the acquisition. In addition, we know from numerical simulations \cite{Job2005} that under perfectly elastic collisions, forces recorded in the chain and at the end of the chain are related in amplitude and duration: $F_m^{wall}\simeq1.94\times F_m^{chain}$ and $\tau_{wall}\simeq1.09\times\tau_{chain}$, with relative error less than $5$~\%.

In Fig.~\ref{fig:Force_vs_position_q1} the evolution of an unaveraged pulse, initiated in a $n_0=16$ beads long monodisperse chain, as it propagates in the first tapered chain ($q=q_1$) made with $n_1=14$ beads of decreasing diameters as described above. Practically, the evolution of the pulse is measured at the end of a tapered chain containing an increasing number of beads ($0\leq n_1\leq14$). It appears that a tail is formed behind a front pulse and the wave, initially symmetric, becomes more and more asymmetric. In Fig.~\ref{fig:Force_vs_position_q2} is the evolution of a pulse, initiated in a $n_0=16$ beads monodisperse chain, as it propagates in the second tapered chain ($q=q_2$) made with $n_2=12$ beads of decreasing diameters as described above. The tail formed behind a front pulse of the wave is in this case even more pronounced. This effect is exactly the one predicted by Sen {\it et al} \cite{Sen2003}. For tapered chains ($q\neq0$), the pulse becomes distributed throughout the chain as it propagates. Since impulsion is transmitted down the chain at the contact surface, impedance mismatches are introduced by tapering the diameters, every mismatch limiting transmission between the grains. By comparing Fig.~\ref{fig:Force_vs_position_q1} to Fig.~\ref{fig:Force_vs_position_q2}, it can be observed that the effect of pulse spreading and maximum force decrease become more dramatic as the tapering factor is increased. Fig.~\ref{fig:Force_vs_position_q1}$h$ and Fig.~\ref{fig:Force_vs_position_q2}$h$ illustrate typical power spectrum of the incident solitary wave, as well as the corresponding waves reaching the end of the chain for the values of tapering factor used here. A clear cascade of energy is observed, from low to high frequencies, as the pulse propagates through the tapered chain.

%--------------------------------------------------------------------%
%-----> FIGURE #3 OVER 6
%--------------------------------------------------------------------%

\begin{figure}[ht]
\includegraphics{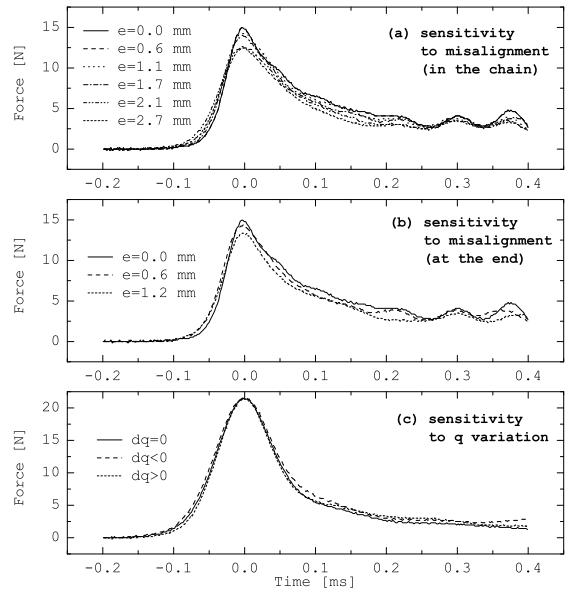}
\caption{\label{fig:Experimental_Sensitivity} Force as a function of time, measured at the end of the entire chain of of tapered factor $q=q_2$. Vertical displacement $e$ of the bead under scope is indicated on the figure. (a) Effect of a controlled vertical misalignment introduced at the bead number $\phi_{22}=16$~mm close to the middle of the tapered chain. (b) Effect of similar vertical misalignment which is this time located at the bead in contact with the wall sensor, $\phi_{28}=9$~mm. (c) The force as a function of time, measured at the end of a partial tapered chain ($q=q_1$) made of beads $01$ to $24$. $dq/q =+1$ corresponds to the same signal when the bead number $18$ is replaced by a bead of diameter $\phi_{18}=\phi_{19}=23.0$~mm. Thus, two consecutive beads have identical diameter. Similarly, for $dq/q =-1$, the bead number $19$ is replaced by a bead of diameter $\phi_{19}^{'}=\phi_{18}=23.0$~mm.}
\end{figure}

In order to check the robustness of our experimental configuration, we have explored how sensitive is the shock mitigation effect to small misalignment of beads. To quantify the effect of beads misalignments we have performed two test experiments. In a first experiment, we check the chain sensitivity to misalignment occurring in the middle of the tapered chain. Second, we impose a well controlled vertical displacement of the bead in contact with the wall sensor and observe how the pulse is modified. As seen in Fig.~\ref{fig:Experimental_Sensitivity}$a$, as the misalignment in the middle of the tapered chain increases, the signal decreases slightly in amplitude and the oscillations of the tail progressively decrease. When looking carefully, one can see that the more pronounced peak of the tail ($t-T\simeq0.38$~ms) is progressively decreasing in amplitude, suggesting the existence of a correlation between the oscillating tail and a given contact at the chain. Similarly, as observed from Fig.~\ref{fig:Experimental_Sensitivity}$b$, a misalignment up to $10$~\% at the end chain does not introduce dramatic effect on the wave. The amplitude is slightly modified and only some effects take place at the tail of the wave. However, in both cases, the front pulse velocity is nearly unchanged at the experimental resolution achieved here.

Although the tapering factor was not kept exactly constant through the chain in the above description, the effect of shock mitigation predicted by Sen {\it et al} \cite{Sen2001} is well captured in our experimental results. To better characterize the effect of local variation of tapered factor, we have tested a short tapered chain in which we replaced one of the bead by a smaller one, in such a way that two consecutive beads have equal diameter. Thus, a local variation of $q$, $dq/q =+1$ is produced. Similarly, to investigate the effect of a negative variation of $q$, $dq/q=-1$, we replace a bead by one equal to the bigger neighbor. As seen in Fig.~\ref{fig:Experimental_Sensitivity}$c$, two effects are visualized. First, for $dq/q=+1$, the front pulse is slightly steeper and narrower. In contrast, for $dq/q=-1$, a smoother and slightly wider front pulse is produced. These two effects are consistent with results presented in Fig.~\ref{fig:Force_vs_position_q1} and Fig.~\ref{fig:Force_vs_position_q2}.

%--------------------------------------------------------------------%
%-----> III) Discussion
%--------------------------------------------------------------------%

\section{Discussion}

The physical behavior of solitary waves in chains of $N$ identical beads is summarized below. Under elastic deformation, the energy stored at the contact between two elastic bodies submitted to an axial compression corresponds to the Hertz potential \cite{Landau1967}, $U_{H}=(2/5)\kappa\delta^{5/2}$, where $\delta$ is the overlap deformation between bodies, $\kappa^{-1}=(\theta+\theta')(R^{-1}+R'^{-1})^{1/2}$, $\theta=3(1-\nu^2)/(4Y)$, and $R$ and $R'$ are radii of curvature at the contact. $Y$ and $\nu$ are Young's Modulus and Poisson's ratio respectively. Since the force felt at the interface is the derivative of the potential with respect to $\delta$, ($F_{H}=\partial_\delta U_H=\kappa\delta^{3/2}$), the dynamics of the chain of beads is described by the following system of $N$ coupled nonlinear equations,
\begin{equation}\label{Eq:DiscreteEquation}
m\ddot{u}_n=\kappa\left[(u_{n-1}-u_{n})^{3/2}_{+}-(u_{n}-u_{n+1})^{3/2}_{+}\right],
\end{equation}
where overdots denote time derivatives, $m$ is the mass, $u_n$ is the displacement of the center of mass of bead $n$, and the label $+$ on the brackets indicates that the Hertz force is zero when the beads loose contact. Under the long-wavelength approximation, $\lambda\gg R$ (where $\lambda$ is the characteristic wavelength of the perturbation), the continuum limit of Eq.~\ref{Eq:DiscreteEquation} can be obtained by replacing the discrete function $u_{n\pm1}(t)$ by the Taylor expansion of the continuous function $u(x\pm2R,t)$. Keeping terms of up to the fourth order spatial derivatives, Eq.~\ref{Eq:DiscreteEquation} leads to the equation for the strain $\psi=-\partial_x u>0$ \cite{Nesterenko2001},
\begin{equation}\label{Eq:ContinuousEquation}
\ddot{\psi}\simeq c^2\partial_{xx}[\psi^{3/2}+(2/5)R^2\psi^{1/4}\partial_{xx}(\psi^{5/4})],
\end{equation}
which admits an exact periodic solution in the form of a travelling wave, $\psi(\xi=x-vt)$, with speed $v$. This solution is \cite{Nesterenko1984,Lazaridi1985,Gavrilyuk1993,Nesterenko1994,Nesterenko1995}:
\begin{equation}\label{Eq:Solution}
\psi=(5/4)^2(v/c)^4\cos^4[\xi/(R\sqrt{10})],
\end{equation}
where $c=(2R)^{5/4}(\kappa/m)^{1/2}$. Although this solution only satisfies the truncated Eq.~\ref{Eq:DiscreteEquation}, it is well established that one hump ($-\pi/2<\xi/(R\sqrt{10})<\pi/2$) of this periodic function represents a solitary wave solution \cite{Nesterenko1984,Chatterjee1999,Job2005}. Finally, approximating spatial derivative by finite difference, the strain in the chain reads $\psi\simeq\delta/(2R)$. The force felt at the interface, $F\simeq\kappa(2R\psi)^{3/2}$, and the wave velocity $v$ become
\begin{eqnarray}
&F \simeq F_m\cos^6{\left[\frac{x-vt}{R\sqrt{10}}\right]},&
\label{Eq:SolutionForce}\\
&v \simeq \left(\frac{6}{5\pi\rho}\right)^{\frac{1}{2}}\left(\frac{F_m}{\theta^{2}R^{2}}\right)^{\frac{1}{6}}
  \simeq \left(\frac{6}{5\pi\rho}\right)^{\frac{2}{5}}\left(\frac{V_m}{\theta^{2}}
  \right)^{\frac{1}{5}}&.
  \label{Eq:SolutionVelocity}
\end{eqnarray}

The second expression for the wave velocity $v$ is obtained by relating the maximum force $F_m$ to the maximum bead's velocity $V_m$ \cite{Nesterenko1984,Lazaridi1985,Nesterenko2001,Landau1967}. Experimentally, the maximum force $F_m$, the time of flight $T$, and the duration $\tau$, are obtained here from fitting the loading part ($-\tau<t-T<0$) of the measured force as a function of time to the Nesterenko solution $F(t)=F_m\cos^6{[(t-T)/\tau]}$.

It is worth mentioning that Rosenau {\it et al}, while studying the role of nonlinear dispersion in pattern formation, derived a Korteweg-de Vries like equation~\cite{Rosenau1993,Rosenau1994,Rosenau1997}, which is similar to the continuum limit of Eq.~\ref{Eq:DiscreteEquation} described above. Among other solutions such as kinks, solitons with infinite slope, periodic waves, dark solitons with cusp, and other non analytical solutions~\cite{Rosenau1997}, they found a solution of the form of Eq.~\ref{Eq:Solution} with a compact support (thus named compactons): their solution is strictly zero outside a finite range, i.e. for $|\xi/(R\sqrt{10})|>\pi/2$~\cite{Rosenau1997}.

In the following, we discuss what features of the front pulse, travelling in the tapered chain, can be captured using the simple monodisperse solution of solitary waves in conjunction with a ballistic approximation. Indeed, Doney {\it et al} \cite{Doney2005} proposed recently a ballistic approximation, with an heuristic term rendering energy dissipation, to model elastic collision of individual beads. Taking into account a classical restitution coefficient, $\epsilon$, the ratio of consecutive bead's velocity is,
\begin{equation}\label{Eq:BeadVelocity}
\frac{V_m^{n+1}}{V_m^{n}}=\frac{2\epsilon}{1+(1-q)^3}.
\end{equation}

\noindent The restitution coefficient is obtained from force measurements \cite{Job2005} acquired in the monodisperse chain ($q=0$), by using that $\epsilon=(F_m^{n+1}/F_m^{n})^{5/6}\equiv V_m^{n+1}/V_m^{n}$. We thus estimate that $\epsilon\simeq0.985$. The theoretical maximum of the force in the tapered chain is then obtained from Eq.~\ref{Eq:SolutionVelocity},
\begin{equation}\label{Eq:MaxForce}
\frac{F_m^{n+1}}{F_m^{n}}=(1-q)^2[\frac{2\epsilon}{1+(1-q)^3}]^{6/5},
\end{equation}
and the predicted speed of the quasisolitary wave (the front pulse), in the ballistic approximation, reads,
\begin{equation}\label{Eq:SolVel}
\frac{v_{n+1}}{v_{n}}=[\frac{2\epsilon}{1+(1-q)^3}]^{1/5}.
\end{equation}

\begin{figure}[ht]
\includegraphics{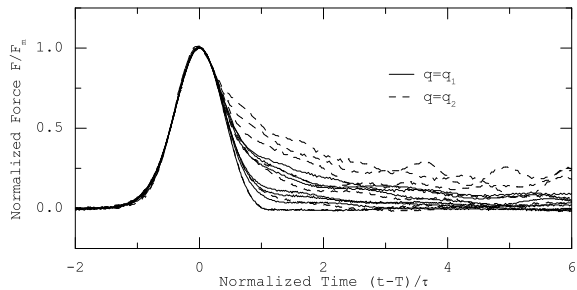}
\caption{\label{fig:Collapse} The collapse of all data presented in Fig.~\ref{fig:Force_vs_position_q1} and Fig.~\ref{fig:Force_vs_position_q2} when the force is normalized to the measured maximum force, $F_m$, and the time scale is $(t-T)/\tau$, where $T$ and $\tau$ are the measured time of flight and pulse duration, respectively.}
\end{figure}

In writing Eq.~\ref{Eq:MaxForce} and Eq.~\ref{Eq:SolVel}, it is assumed that the front pulse adapts itself to the self similar solution at every impedance mismatch. Therefore, this approximation is suitable for small tapering factor, since for slow variation of bead size, quasisolitary wave adjustments are expected to be small and to take place at distance much longer than its characteristic width. This can be shown also by comparing the time scale of the wave, $\tau_{wave}\sim 2R/v$, to the time scale of the loading during a collision, $\tau_{coll}\sim \delta_m/V_m$, (use Eq.~\ref{Eq:SolutionVelocity} to estimate the ratio $v/V_m$). These characteristic times are equal in a monodisperse chain and almost equal for small tapering factor, demonstrating that the loading of the contact is the process governing the main features of quasisolitary wave formation. In Fig.~\ref{fig:Collapse}, when the force is normalized to the maximum force, $F_m$, and the time scale is selected to be the measured duration $\tau$, all data presented in Fig.~\ref{fig:Force_vs_position_q1} and Fig.~\ref{fig:Force_vs_position_q2} show a collapse in the rising part of the front pulse. Thus, this result provides strong evidence that the main front of the pulse follows a self similar solution.

%--------------------------------------------------------------------%
%-----> FIGURE #4 OVER 6
%--------------------------------------------------------------------%

\begin{figure}[ht]
\includegraphics{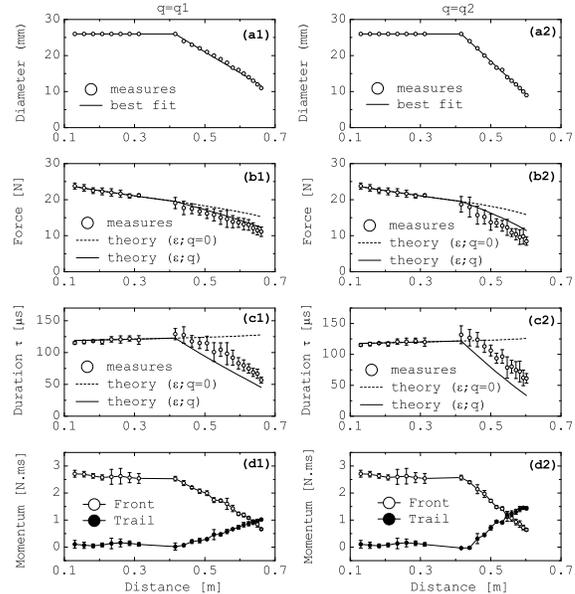}
\caption{\label{fig:Waves_Characteristics}. Characteristics of the chain and of the pulse traveling in both tapered chains (the left column is for $q=q_1$ and the right one is for $q=q_2$) as a function of the position in the chain. (a1)-(a2): the beads diameter used in the experiments compared with the fit presented in the text. (b1)-(b2): maximum compression measured in the chain. (c1)-(c2): duration of the pulse contrasted to predictions from dissipative ballistic approximation including or not the tapering effect. (d1)-(d2): front and tail linear momentum, respectively $Q_{F}=\int_{-\infty}^{T+\tau}{F(t)dt}$ and $Q_{T}=\int_{T+\tau}^{+\infty}{F(t)dt}$. Here the solid lines are guides for the eye.}
\end{figure}

Based on the results above, we discuss in further detail the main features of the quasisolitary wave, namely, its maximum force, its local speed and the momentum transfer from the main peak to its tail. As a reference, for both tapering factors, Fig.~\ref{fig:Waves_Characteristics}$a1$ and Fig.~\ref{fig:Waves_Characteristics}$a2$ indicate the size of beads as a function of position. In Fig.~\ref{fig:Waves_Characteristics}$b1$ and Fig.~\ref{fig:Waves_Characteristics}$b2$ the maximum force felt by the rigid wall. An important decrease in the force is observed in both cases as the pulse propagates in the tapered chain. Interestingly, such a decrease can not be attributed to the dissipation only. Indeed, dashed line in Fig.~\ref{fig:Waves_Characteristics}$b1$ and Fig.~\ref{fig:Waves_Characteristics}$b2$ indicate that predictions from Eq.~\ref{Eq:MaxForce}, with $q=0$, overestimate the force. However, when tapering factor is introduced, the experimental data follow reasonably well the predictions from Eq.~\ref{Eq:MaxForce}.

%--------------------------------------------------------------------%
%-----> FIGURE #5 OVER 6
%--------------------------------------------------------------------%

In Fig.~\ref{fig:Waves_Characteristics}$c1$ and Fig.~\ref{fig:Waves_Characteristics}$c2$ we represent the duration $\tau$ of the pulses as the bead diameter decreases. From the theory of monodisperse chain schematized above, it is known that the quasisolitary wave duration $\tau$, according to Eq.~\ref{Eq:SolutionForce}, is $\tau_n=(\phi_n/2)\sqrt{10}/v_n$. It has been checked that $\tau_n$, measured at every bead, agrees with the monodisperse-ballistic approximation in a reasonable manner, for small tapering factor, see solid lines on Fig.~\ref{fig:Waves_Characteristics}$c1$ and Fig.~\ref{fig:Waves_Characteristics}$c2$. This result is not obvious since to calculate the solid line on Fig.~\ref{fig:Waves_Characteristics}$c1$ and Fig.~\ref{fig:Waves_Characteristics}$c2$, we have used the calculated value of the quasisolitary wave speed at every bead from Eq.~\ref{Eq:SolVel}.

On the other hand, the impulse of the front is $Q_{F}=\int_{-\infty}^{T+\tau}{F(t)dt}$ and the impulse of the trail is $Q_{T}=\int_{T+\tau}^{+\infty}{F(t)dt}$. The principal effect of the tapered chain is then shown in Fig.~\ref{fig:Waves_Characteristics}$d1$ and Fig.~\ref{fig:Waves_Characteristics}$d2$, in which a noticeable part of the impulse is transferred from the front pulse to the tail. From comparison to Fig.~\ref{fig:Waves_Characteristics}$b$ and Fig.~\ref{fig:Waves_Characteristics}$c$, it appears that the diminution of the front impulse is principally due to both dramatic decreases of the maximum force and of the duration of the front pulse, as it propagates down the tapered chain. Note that the total momentum is conserved in the tapered chain. This has been checked more accurately using lower frequency resolution, which allowed us to fully record the pulse tail and thus verify that ($Q_F+Q_T=cte$). In fact, the signal being acquired on a fixed set of points, for the rest of our measurements, we have preferred to keep a high sampling frequency, which provides high resolution for the front pulse but avoids to record the very end of its tail.

%--------------------------------------------------------------------%
%-----> FIGURE #6 OVER 6
%--------------------------------------------------------------------%

\begin{figure}[ht]
\includegraphics{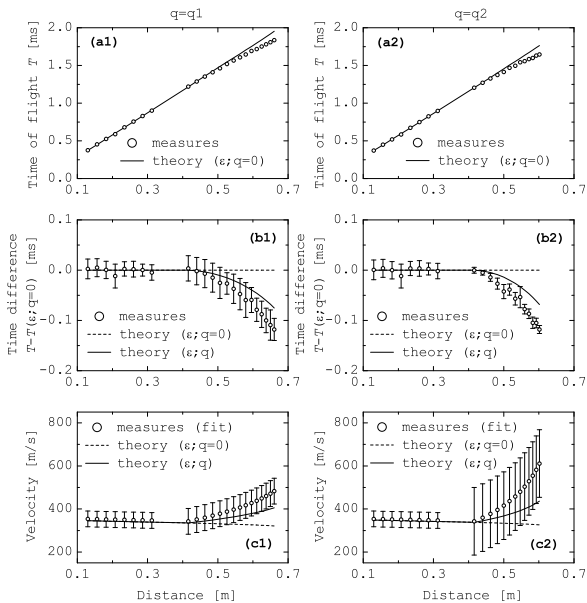}
\caption{\label{fig:Time_Characteristics} Time of flight and velocity measurements, in both tapered chains, as a function of the position in the chain. (a1)-(a2): flight time $T$. (b1)-(b2): difference between the measured flight time and the theoretical flight of time obtained when only the restitution coefficient is taken into account ($\epsilon<1$ and $q=0$). (c1)-(c2): pulse velocity.}
\end{figure}

We compare below the measured to the predicted time of flight of the front pulse when reaching the sensing wall. The relation $T_n=T_0+\sum_{1}^{n}{(\phi_m/v_m)}=T_0+(\phi_0/v_1)\frac{1-\epsilon^{-n/5}}{1-\epsilon^{-1/5}}$ helps to find $v_1$ and $T_0$, by fitting to measurements done in a chain of equal beads ($n\leq n_0$). The theoretical velocity is then found by using Eq.~\ref{Eq:SolVel}, and finally, a prediction of the time of flight in the tapered chain is obtained by using the relation $T_{n}=T_{n-1}+(\phi_{n}/v_{n})$. In Fig.~\ref{fig:Time_Characteristics}$a1$ and Fig.~\ref{fig:Time_Characteristics}$a2$ comparisons with experimental data are presented in the case of $q=0$. The decrease observed for both tapered factors can not be attributed to the restitution coefficient effect only. The effect of tapering factor is better observed in Fig.~\ref{fig:Time_Characteristics}$b1$ and Fig.~\ref{fig:Time_Characteristics}$b2$ which provide a close comparison of time difference, $T-T(\varepsilon; q=0)$, where $T(\varepsilon;q=0)$ is the prediction that only includes the restitution coefficient effect. It is seen that the measured time of flight is shorter than the prediction from ballistic approximation which slightly overestimates the time of flight between two consecutive beads.

For the sake of comparison, we finally contrast theoretical velocity from Eq.~\ref{Eq:SolVel} to local velocity $v_n^d=(\delta_n-\delta_{n-1})/(T_n-T_{n-1})$. Experimentally, the local velocity, estimated from time of flight differences, naturally provides very large relative errors. We thus choose to present an estimate obtained by smoothing the local time of flight fluctuations. This is done as follow. In accordance to Eq.~\ref{Eq:SolVel}, we map the local velocity $v_n^d$ to the behavior $v_n^d=Q^{n-n'}v_{n'}^d$ to find equivalent coefficients of proportionality $Q$ for both tapered chains. Results of the fits are presented in Fig.~\ref{fig:Time_Characteristics}$c1$ and Fig.~\ref{fig:Time_Characteristics}$c2$. The inflection of the velocity in the tapered part of the chain, rendering the acceleration of the pulse, is clearly demonstrated in both cases. Nevertheless, experimental fits slightly overestimate theoretical predictions, and a closer comparison is necessary. Experimentally, we find that the coefficient of proportionality is $Q_1=1.025\pm0.016$ for the first chain ($q=q_1$), and $Q_2=1.049\pm0.044$ for the second chain ($q=q_2$). These coefficients must be compared to their theoretical predictions, respectively $Q_1^{th}=1.014$ and $Q_2^{th}=1.022$. We thus check that both theoretical values lie inside their respective experimental errorbars. Moreover these experimental coefficients only overestimate their theoretical counterparts by $1.1$~\% for the first chain, and by $2.7$~\% for the second one. This last satisfactory agreement, between theoretical and experimental coefficients of proportionality $Q$ in both chains, is not surprising and must be balanced, since $Q$ is weakly sensitive to tapering factor variation $\delta q$ (see Eq.~\ref{Eq:SolVel} with $q<<1$): $\delta Q/Q=(3/10)\delta q$.

%--------------------------------------------------------------------%
%-----> IV) Conclusion
%--------------------------------------------------------------------%

\section{Conclusion}

In conclusion, we have shown that a tapered chain is useful for shock attenuation and energy transfer from low to high frequencies, as previously stated by Sen {\it et al} \cite{Sen2001,Sen2003,Doney2005,Nakagawa2004,SenNaka2005}. The ballistic approximation allowed capturing essential behaviors such as front pulse acceleration, amplitude shrinking, and diminution of its duration. This lowest order model appeared accurate for small tapering factor, but may become insufficient when tapering becomes greater. Under this approximation, it was assumed that the front pulse adapts itself to a self similar solution at every impedance mismatch. This may clearly not always be true. Nevertheless, our data seemed to follow this scaling in a reasonable good manner indicating that the main parameters to characterize the quasisolitary wave are the maximum force and the beads size, in addition to the mechanical properties of the beads. As a possible extension of this work, it would be interesting to extract a more complete behavior from a nonlinear wave equation that would take into account the tapering of the chain. In addition, comparisons of our experimental findings with numerical simulations are under way and should help to elucidate some of the hypothesis stated here.

%--------------------------------------------------------------------%
%-----> Acknowledgments
%--------------------------------------------------------------------%

\begin{acknowledgments}
This work was supported by Conicyt-Chile under research program Fondap $N^{o}$ 11980002.
\end{acknowledgments}

%--------------------------------------------------------------------%
%-----> Bibliography
%--------------------------------------------------------------------%

\bibliography{Exp_Tapered_Chain_Melo_Job}

\begin{thebibliography}{27}
\expandafter\ifx\csname natexlab\endcsname\relax\def\natexlab#1{#1}\fi
\expandafter\ifx\csname bibnamefont\endcsname\relax
  \def\bibnamefont#1{#1}\fi
\expandafter\ifx\csname bibfnamefont\endcsname\relax
  \def\bibfnamefont#1{#1}\fi
\expandafter\ifx\csname citenamefont\endcsname\relax
  \def\citenamefont#1{#1}\fi
\expandafter\ifx\csname url\endcsname\relax
  \def\url#1{\texttt{#1}}\fi
\expandafter\ifx\csname urlprefix\endcsname\relax\def\urlprefix{URL }\fi
\providecommand{\bibinfo}[2]{#2}
\providecommand{\eprint}[2][]{\url{#2}}

\bibitem[{\citenamefont{Nesterenko}(2001)}]{Nesterenko2001}
\bibinfo{author}{\bibfnamefont{V.~F.} \bibnamefont{Nesterenko}},
  \emph{\bibinfo{title}{Dynamics of heterogeneous materials}}
  (\bibinfo{publisher}{Springer-Verlag}, \bibinfo{address}{New York},
  \bibinfo{year}{2001}).

\bibitem[{\citenamefont{S.~Sen and Manciu}(2001)}]{Sen2001}
\bibinfo{author}{\bibfnamefont{F.~S.~M.} \bibnamefont{S.~Sen}}
  \bibnamefont{and} \bibinfo{author}{\bibfnamefont{M.}~\bibnamefont{Manciu}},
  \bibinfo{journal}{Physica A} \textbf{\bibinfo{volume}{299}},
  \bibinfo{pages}{551} (\bibinfo{year}{2001}).

\bibitem[{\citenamefont{Sen et~al.}(2003)\citenamefont{Sen, Chakravarti, Visco,
  Wu, Nakagawa, and Agui}}]{Sen2003}
\bibinfo{author}{\bibfnamefont{S.}~\bibnamefont{Sen}},
  \bibinfo{author}{\bibfnamefont{S.}~\bibnamefont{Chakravarti}},
  \bibinfo{author}{\bibfnamefont{D.~P.} \bibnamefont{Visco}},
  \bibinfo{author}{\bibfnamefont{D.~T.} \bibnamefont{Wu}},
  \bibinfo{author}{\bibfnamefont{M.}~\bibnamefont{Nakagawa}}, \bibnamefont{and}
  \bibinfo{author}{\bibfnamefont{J.~H.} \bibnamefont{Agui}}, in
  \emph{\bibinfo{booktitle}{AIP Conf. Proc.}}, edited by
  \bibinfo{editor}{\bibfnamefont{K.}~\bibnamefont{Lindenberg}}
  (\bibinfo{year}{2003}), vol. \bibinfo{volume}{658}, p. \bibinfo{pages}{357}.

\bibitem[{\citenamefont{Doney and Sen}(2005)}]{Doney2005}
\bibinfo{author}{\bibfnamefont{R.~L.} \bibnamefont{Doney}} \bibnamefont{and}
  \bibinfo{author}{\bibfnamefont{S.}~\bibnamefont{Sen}},
  \bibinfo{journal}{Phys. Rev. E} \textbf{\bibinfo{volume}{72}},
  \bibinfo{pages}{041304} (\bibinfo{year}{2005}).

\bibitem[{\citenamefont{Sokolow et~al.}(2005)\citenamefont{Sokolow, Pfannes,
  Doney, Nakagawa, Agui, and Sen}}]{SenNaka2005}
\bibinfo{author}{\bibfnamefont{A.}~\bibnamefont{Sokolow}},
  \bibinfo{author}{\bibfnamefont{J.~M.} \bibnamefont{Pfannes}},
  \bibinfo{author}{\bibfnamefont{R.~L.} \bibnamefont{Doney}},
  \bibinfo{author}{\bibfnamefont{M.}~\bibnamefont{Nakagawa}},
  \bibinfo{author}{\bibfnamefont{J.~H.} \bibnamefont{Agui}}, \bibnamefont{and}
  \bibinfo{author}{\bibfnamefont{S.}~\bibnamefont{Sen}},
  \bibinfo{journal}{Applied Physics Letters} \textbf{\bibinfo{volume}{87}}
  (\bibinfo{year}{2005}).

\bibitem[{\citenamefont{Hong}(2005)}]{Hong2005}
\bibinfo{author}{\bibfnamefont{J.}~\bibnamefont{Hong}}, \bibinfo{journal}{Phys.
  Rev. Lett.} \textbf{\bibinfo{volume}{94}}, \bibinfo{pages}{108001}
  (\bibinfo{year}{2005}).

\bibitem[{\citenamefont{Nesterenko}(1984)}]{Nesterenko1984}
\bibinfo{author}{\bibfnamefont{V.~F.} \bibnamefont{Nesterenko}},
  \bibinfo{journal}{J. Appl. Mech. Tech. Phys.} \textbf{\bibinfo{volume}{24}},
  \bibinfo{pages}{733} (\bibinfo{year}{1984}).

\bibitem[{\citenamefont{Lazaridi and Nesterenko}(1985)}]{Lazaridi1985}
\bibinfo{author}{\bibfnamefont{A.~N.} \bibnamefont{Lazaridi}} \bibnamefont{and}
  \bibinfo{author}{\bibfnamefont{V.~F.} \bibnamefont{Nesterenko}},
  \bibinfo{journal}{J. Appl. Mech. Tech. Phys.} \textbf{\bibinfo{volume}{26}},
  \bibinfo{pages}{405} (\bibinfo{year}{1985}).

\bibitem[{\citenamefont{Nesterenko}(1994)}]{Nesterenko1994}
\bibinfo{author}{\bibfnamefont{V.~F.} \bibnamefont{Nesterenko}},
  \bibinfo{journal}{J. Phys. IV} \textbf{\bibinfo{volume}{4}},
  \bibinfo{pages}{C8} (\bibinfo{year}{1994}).

\bibitem[{\citenamefont{Nesterenko et~al.}(1995)\citenamefont{Nesterenko,
  Lazaridi, and Sibiryakov}}]{Nesterenko1995}
\bibinfo{author}{\bibfnamefont{V.~F.} \bibnamefont{Nesterenko}},
  \bibinfo{author}{\bibfnamefont{A.~N.} \bibnamefont{Lazaridi}},
  \bibnamefont{and} \bibinfo{author}{\bibfnamefont{E.~B.}
  \bibnamefont{Sibiryakov}}, \bibinfo{journal}{J. Appl. Mech. Tech. Phys.}
  \textbf{\bibinfo{volume}{36}}, \bibinfo{pages}{166} (\bibinfo{year}{1995}).

\bibitem[{\citenamefont{Coste et~al.}(1997)\citenamefont{Coste, Falcon, and
  Fauve}}]{Coste1997}
\bibinfo{author}{\bibfnamefont{C.}~\bibnamefont{Coste}},
  \bibinfo{author}{\bibfnamefont{E.}~\bibnamefont{Falcon}}, \bibnamefont{and}
  \bibinfo{author}{\bibfnamefont{S.}~\bibnamefont{Fauve}},
  \bibinfo{journal}{Phys. Rev. E} \textbf{\bibinfo{volume}{56}},
  \bibinfo{pages}{6104} (\bibinfo{year}{1997}).

\bibitem[{\citenamefont{Hascoet et~al.}(1999)\citenamefont{Hascoet, Herrmann,
  and Loreto}}]{Hascoet1999}
\bibinfo{author}{\bibfnamefont{E.}~\bibnamefont{Hascoet}},
  \bibinfo{author}{\bibfnamefont{H.~J.} \bibnamefont{Herrmann}},
  \bibnamefont{and} \bibinfo{author}{\bibfnamefont{V.}~\bibnamefont{Loreto}},
  \bibinfo{journal}{Phys. Rev. E} \textbf{\bibinfo{volume}{59}},
  \bibinfo{pages}{3202} (\bibinfo{year}{1999}).

\bibitem[{\citenamefont{Mackay}(1999)}]{Mackay99}
\bibinfo{author}{\bibfnamefont{R.~S.} \bibnamefont{Mackay}},
  \bibinfo{journal}{Phys. Lett. A} \textbf{\bibinfo{volume}{251}},
  \bibinfo{pages}{191} (\bibinfo{year}{1999}).

\bibitem[{\citenamefont{Job et~al.}(2005)\citenamefont{Job, Melo, Sokolow, and
  Sen}}]{Job2005}
\bibinfo{author}{\bibfnamefont{S.}~\bibnamefont{Job}},
  \bibinfo{author}{\bibfnamefont{F.}~\bibnamefont{Melo}},
  \bibinfo{author}{\bibfnamefont{A.}~\bibnamefont{Sokolow}}, \bibnamefont{and}
  \bibinfo{author}{\bibfnamefont{S.}~\bibnamefont{Sen}},
  \bibinfo{journal}{Phys. Rev. Lett.} \textbf{\bibinfo{volume}{94}},
  \bibinfo{pages}{178002} (\bibinfo{year}{2005}).

\bibitem[{\citenamefont{P{\"o}schel and Brilliantov}(2001)}]{Poschel2001}
\bibinfo{author}{\bibfnamefont{T.}~\bibnamefont{P{\"o}schel}} \bibnamefont{and}
  \bibinfo{author}{\bibfnamefont{N.~V.} \bibnamefont{Brilliantov}},
  \bibinfo{journal}{Phys. Rev. E} \textbf{\bibinfo{volume}{63}},
  \bibinfo{pages}{021 505} (\bibinfo{year}{2001}).

\bibitem[{\citenamefont{Sinkovits and Sen}(1995)}]{Sinkovits95}
\bibinfo{author}{\bibfnamefont{R.~S.} \bibnamefont{Sinkovits}}
  \bibnamefont{and} \bibinfo{author}{\bibfnamefont{S.}~\bibnamefont{Sen}},
  \bibinfo{journal}{Phys. Rev. Lett.} \textbf{\bibinfo{volume}{74}},
  \bibinfo{pages}{2686} (\bibinfo{year}{1995}).

\bibitem[{\citenamefont{Sen and Sinkovits}(1996)}]{Sen96}
\bibinfo{author}{\bibfnamefont{S.}~\bibnamefont{Sen}} \bibnamefont{and}
  \bibinfo{author}{\bibfnamefont{R.~S.} \bibnamefont{Sinkovits}},
  \bibinfo{journal}{Phys. Rev. E} \textbf{\bibinfo{volume}{54}},
  \bibinfo{pages}{6857} (\bibinfo{year}{1996}).

\bibitem[{\citenamefont{Hong et~al.}(1999)\citenamefont{Hong, Ji, and
  Kim}}]{Hong99}
\bibinfo{author}{\bibfnamefont{J.}~\bibnamefont{Hong}},
  \bibinfo{author}{\bibfnamefont{J.-Y.} \bibnamefont{Ji}}, \bibnamefont{and}
  \bibinfo{author}{\bibfnamefont{H.}~\bibnamefont{Kim}},
  \bibinfo{journal}{Phys. Rev. Lett.} \textbf{\bibinfo{volume}{82}},
  \bibinfo{pages}{3058} (\bibinfo{year}{1999}).

\bibitem[{\citenamefont{Hascoet and Hinch}(2002)}]{Hascoet02}
\bibinfo{author}{\bibfnamefont{E.}~\bibnamefont{Hascoet}} \bibnamefont{and}
  \bibinfo{author}{\bibfnamefont{E.~J.} \bibnamefont{Hinch}},
  \bibinfo{journal}{Phys. Rev. E} \textbf{\bibinfo{volume}{66}},
  \bibinfo{pages}{011307} (\bibinfo{year}{2002}).

\bibitem[{\citenamefont{Nakagawa et~al.}(2004)\citenamefont{Nakagawa, Agui, Wu,
  and Extramiana}}]{Nakagawa2004}
\bibinfo{author}{\bibfnamefont{M.}~\bibnamefont{Nakagawa}},
  \bibinfo{author}{\bibfnamefont{J.~H.} \bibnamefont{Agui}},
  \bibinfo{author}{\bibfnamefont{D.~T.} \bibnamefont{Wu}}, \bibnamefont{and}
  \bibinfo{author}{\bibfnamefont{D.~V.} \bibnamefont{Extramiana}},
  \bibinfo{journal}{Gran. Mat.} \textbf{\bibinfo{volume}{4}},
  \bibinfo{pages}{167} (\bibinfo{year}{2004}).

\bibitem[{Tsu()}]{Tsubaki}
\bibinfo{note}{See for instance, http://www.wsb.co.th/.}

\bibitem[{\citenamefont{Landau and Lifshitz}(1967)}]{Landau1967}
\bibinfo{author}{\bibfnamefont{L.~D.} \bibnamefont{Landau}} \bibnamefont{and}
  \bibinfo{author}{\bibfnamefont{E.~M.} \bibnamefont{Lifshitz}},
  \emph{\bibinfo{title}{Theorie de l'\'elasticit\'e}}
  (\bibinfo{publisher}{Mir}, \bibinfo{address}{Moscou}, \bibinfo{year}{1967}),
  \bibinfo{edition}{2nd} ed., \bibinfo{note}{in French}.

\bibitem[{\citenamefont{Gavrilyuk and Nesterenko}(1993)}]{Gavrilyuk1993}
\bibinfo{author}{\bibfnamefont{S.~L.} \bibnamefont{Gavrilyuk}}
  \bibnamefont{and} \bibinfo{author}{\bibfnamefont{V.~F.}
  \bibnamefont{Nesterenko}}, \bibinfo{journal}{J. Appl. Mech. Tech. Phys.}
  \textbf{\bibinfo{volume}{34}}, \bibinfo{pages}{784} (\bibinfo{year}{1993}).

\bibitem[{\citenamefont{Chatterjee}(1999)}]{Chatterjee1999}
\bibinfo{author}{\bibfnamefont{A.}~\bibnamefont{Chatterjee}},
  \bibinfo{journal}{Phys. Rev. E} \textbf{\bibinfo{volume}{59}},
  \bibinfo{pages}{5912} (\bibinfo{year}{1999}).

\bibitem[{\citenamefont{Rosenau and Hyman}(1993)}]{Rosenau1993}
\bibinfo{author}{\bibfnamefont{P.}~\bibnamefont{Rosenau}} \bibnamefont{and}
  \bibinfo{author}{\bibfnamefont{J.~M.} \bibnamefont{Hyman}},
  \bibinfo{journal}{Phys. Rev. Lett.} \textbf{\bibinfo{volume}{70}},
  \bibinfo{pages}{564} (\bibinfo{year}{1993}).

\bibitem[{\citenamefont{Rosenau}(1994)}]{Rosenau1994}
\bibinfo{author}{\bibfnamefont{P.}~\bibnamefont{Rosenau}},
  \bibinfo{journal}{Phys. Rev. Lett.} \textbf{\bibinfo{volume}{73}},
  \bibinfo{pages}{1737} (\bibinfo{year}{1994}).

\bibitem[{\citenamefont{Rosenau}(1997)}]{Rosenau1997}
\bibinfo{author}{\bibfnamefont{P.}~\bibnamefont{Rosenau}},
  \bibinfo{journal}{Phys. Lett. A} \textbf{\bibinfo{volume}{230}},
  \bibinfo{pages}{305} (\bibinfo{year}{1997}).

\end{thebibliography}

%--------------------------------------------------------------------%
%-----> End of Document
%--------------------------------------------------------------------%

\end{document}